\documentclass[11pt,twocolumn,twoside,a4paper,amsmath,amssymb,aps,showkeys,showpacs]{revtex4}
\usepackage{amsfonts}
\usepackage{graphics} %   <------- for LaTeX + EPS
\usepackage{epsfig}
\usepackage{fancyheadings}

\def\simgr{^>\hskip -2.5mm_\sim}               %  greater than or approx equal
\newcommand{\JETSET}{{\scshape Jetset}}
\newcommand{\GeV}{\mathrm{GeV}}
\newcommand{\Lthree}{{\scshape l}{\small 3}\ }
\newcommand{\Pep}{\ensuremath{e^+}}
\newcommand{\Pem}{\ensuremath{e^-}}
\newcommand{\ycut}{\ensuremath{y_\mathrm{cut}}}%             % ycut
\newcommand{\pt}{\ensuremath{p_\mathrm{t}}}%               % pt
\newcommand{\mt}{\ensuremath{m_\mathrm{t}}}%                 % mt
\newcommand{\taumodel}{$\tau$-model}
 
\newcommand{\chisq}{\ensuremath{\chi^2}}%                    % chisq
 
\newcommand{\Eq}[1]{Eq.\,(\ref{#1})}%
\newcommand{\Fig}[1]{Fig.\,\ref{#1}}%
\newcommand{\Tab}[1]{Table~\ref{#1}}%
\newcommand{\ee}{\ensuremath{\mathrm{e^{+}e^{-}}}}

\newcommand{\invGeV}{\GeV\ensuremath{^{-1}}}

\newcommand{\pho}{\phantom{0}}

\begin{document}
\thispagestyle{myheadings}
%%%%%%%%%%%%%%%%%%%%%%%%%% Title %%%%%%%%%%%%%%%%%%%%%%%%%%%%%%%%%%%%%%
\rhead[]{}%<------
\lhead[]{}%<------
\chead[T. Cs\"org\H{o} et al. for \Lthree]{Bose-Einstein correlations at LEP \Lthree}%<------short title

\title{
Detailed L3 measurements of Bose-Einstein correlations and \\
a region of  anti-correlations in hadronic $Z^0$ decays at LEP}

\author{T. Cs\"org\H{o}}
\affiliation{Department of Physics, Harvard University, 17 Oxford St, Cambridge, MA 02138, USA}
\email{csorgo@physics.harvard.edu}
\affiliation{MTA KFKI RMKI, H-1525 Budapest 114, P.O.Box 49, Hungary}

\author{W. Metzger}

%\email{wes@hef.kun.nl}

\affiliation{%
Dept. Experimental High Energy Physics, Radboud University Nijmegen,
P.O. Box 9010,
6500 GL Nijmegen,
The Netherlands
}%

\author{T. Nov\'ak}
%\email{novakt@rmki.kfki.hu}

\affiliation{MTA KFKI RMKI, H-1525 Budapest 114, P.O.Box 49, Hungary
}

\altaffiliation[Presently at ]{Dept. of Business Mathematics \& Informatics, K\'aroly R\'obert College, H-3200 Gy\"ongy\"os, Hungary}

\author{W. Kittel}
%\altaffiliation[Also at ]{JIPNR, Minsk, Belarus}
%\email{wolfram@hef.kun.nl}

\affiliation{%
Dept. Experimental High Energy Physics, Radboud University Nijmegen,
P.O. Box 9010,
6500 GL Nijmegen,
The Netherlands
}%

\author{for the \Lthree Collaboration}
\noaffiliation

\received{February 3, 2010 }

\begin{abstract}
\Lthree preliminary data of two-particle Bose-Einstein correlations
are reported for hadronic $Z^0$ decays in \ee annihilation at LEP. 
The invariant relative momentum $Q$ is identified as the eigenvariable of the 
measured correlation function.  Significant anti-correlations are observed in the Bose-Einstein
correlation function in a broad region of $0.5 - 1.6 $ GeV with a minimum at $Q \approx 0.8 $ GeV.
Absence of Bose-Einstein correlations is demonstrated
in the region above $Q \simgr 1.6$ GeV.
The effective source size is found to decrease 
with increasing value of the transverse mass of the pair,
similarly to hadron-hadron and heavy ion reactions. 
These feautes and our data are 
described well by the non-thermal  $\tau$-model,  which is
based on strong space-time momentum-correlations.
\end{abstract}

\pacs{13.38.Dg, 13.66.Bc, 13.66.Jn, 25.75.Gz}

\keywords{Femtoscopic correlations, Bose-Einstein correlations,
single particle spectra, source function reconstruction, 
$e^+ e^-$ annihilation, hadronic $Z^0$ decays}

\maketitle
 
\clearpage
%*****************   The Body of the Article:   *************************
%\section{Introduction}
%\label{sec:Introduction}

{\it Introduction:} Boson interferometry provides a powerful femtoscopic tool for the
investigation of the space-time structure of particle production processes
on the femtometer lengthscales.  Bose-Einstein correlations (BEC) of two identical 
bosons reflect both geometrical and dynamical properties of 
the particle radiating 
source~\cite{Gyulassy:1979yi,Csorgo:1995bi,Wiedemann:1999qn,Weiner:1999th,Csorgo:1999sj,Lisa:2005dd,Csorgo:2005gd}.
 
In \ee\  annihilation BEC have been observed~\cite{Althoff:1986wn}
to be maximal when the invariant momentum difference of the bosons,
$Q=\sqrt{-(p_1-p_2)^2}$, is small,
even when one of the relative momentum components is large.
This is not the case in hadron-hadron interactions~\cite{Agababyan:1996rg}
or in heavy-ion interactions~\cite{Ahle:2002mi},
where BEC are found not to depend simply on $Q$, but to
decrease even if $Q$ is small but any of the relative momentum components 
is large. 
The size (radius) of the source in heavy-ion collisions
has been found to decrease with increasing transverse momentum, \pt,
or transverse mass, $\mt=\sqrt{m^2+\pt^2}$, % =\sqrt{E^2-p_z^2}$,
of the bosons, for a recent on data oriented review see ref.~\cite{Lisa:2005dd}.  
A similar effect has been seen in $p + p$ collisions~\cite{Chajecki:2005zw}, 
as well as in \ee\ annihilation~\cite{collaboration:2007he}.
Such a behavior that can be described by hydrodynamical models of the source,
~\cite{Csorgo:1999sj}, however, a simple $Q$ dependence of
Bose-Einstein correlations in \ee collisions  
can not be explained in hydrodynamical or thermal models~\cite{Csorgo:2009dn}.

{\it Event and track selection:} 
The data used in the present  analysis were collected by the
\Lthree\ detector at LEP at an \ee 
center-of-mass energy of $\sqrt{s}\approx 91.2$ GeV.
In total about 0.8 million events with an average number of about 12 well-measured charged tracks are
selected. This results in approximately 36 million like-sign pairs of well-measured charged tracks.
Events are classified as two- or three-jet events on the basis of the
Durham jet algorithm with a jet resolution parameter $\ycut=0.006$, 
yielding about 0.5 million two-jet and 0.3 million three-jet events.  
In the present study we report only about the results for two-jet events,
more detailed three-jet results will be presented elsewhere.

{\it Bose-Einstein correlation function:} 
The two-particle correlation function of two particles with
four-momenta $p_{1}$ and $p_{2}$ is given by the ratio of the two-particle number density,
$\rho_2(p_{1},p_{2})$,
to the product of the two single-particle number densities, $\rho_1 (p_{1})\rho_1 (p_{2})$.
Since we are here interested only in the correlation $R_2$ due to Bose-Einstein
interference, the product of single-particle densities is replaced by
$\rho_0(p_1,p_2)$,
the two-particle density that would occur in the absence of Bose-Einstein correlations:
 
\begin{equation} \label{eq:R2def}
  R_2(p_1,p_2)=\frac{\rho_2(p_1,p_2)}{\rho_0(p_1,p_2)} \;.
\end{equation}
An event mixing technique is used to construct $\rho_0$, whereby all tracks of each data
event are replaced by  tracks from  different events having similar multiplicity to the original event.

$\rho_0$ is corrected for detector acceptance and efficiency in the same way as $\rho_2$.
The mixing technique removes all correlations, for example, resonances and energy-momentum conservation,
not just Bose-Einstein correlations.
Hence, $\rho_0$  is also corrected for this 
by a multiplicative factor which is the ratio of the densities of events to mixed events found using
events generated by \JETSET~\cite{Sjostrand:1993yb}, without BEC simulation.
Thus $R_2$ is measured by
\begin{equation} \label{eq:R2cor}
  R_2 = \left(R_\textrm{2\;data} R_\textrm{2\;gen}\right) / \left(R_\textrm{2\;det} R_\textrm{2\;gen-noBE}\right) \;,
\end{equation}
where data, gen, det, gen-noBE refer, respectively, to the data sample, a generator-level Monte Carlo sample,
the same Monte Carlo sample passed through detector simulation and subjected to the same selection procedure as the data,
and a generator-level sample of a Monte Carlo without BEC simulation.

{\it The invariant relative momentum $Q$ is eigenvariable of the correlation function in \ee annihilation. }
%\label{sec:Qinv}
In \ee\ annihilation at lower energy~\cite{Althoff:1986wn} it has been observed that $Q$ is the
appropriate (eigen)variable of the Bose-Einstein correlation function, which implies an approximate
spherical symmetry of particle emission in the rest frame of the pair.
A priori, one does not expect the hadron source to be so spherically symmetric in jet fragmentation.
Recent investigations have, in fact, found an elongation of the source along the
jet axis~\cite{Acciarri:1999ry,Abbiendi:2000gc,Abreu:1999xf,Heister:2003ai,collaboration:2007he}.
While this effect is      well established, the elongation is actually only about 20\%,
which suggests that a parametrization in terms of the single variable $Q$,
may be a good approximation.

We have checked on \Lthree data, if indeed $Q$ is an eigenvariable of the BEC or not, and confirmed 
\cite{Novak:2008zz} that this is indeed the case, both for all and for two-jet events:
We observe that $R_2$ does not decrease when both $q^2=(\vec{p}_1-\vec{p}_2)^2$ and
$q_0^2=(E_1-E_2)^2$ are large while $Q^2=q^2-q_0^2$ is small, but is maximal for
$Q^2=q^2-q_0^2=0$,
independent of the individual values of $q$ and $q_0$.
Furthermore, two-dimensional fits with the parametrization
\begin{equation}  \label{eq:R2qq0}
   R_2(q, q_0) = 1 + \lambda \exp \left((rq)^2-(r_0q_0)^2\right)
\end{equation}
find $r$ and $r_0$ to be equal. (Where we also note parameter $\lambda$, 
the intercept parameter of the correlation  function.)
The similar conclusion is found in a different decomposition: $Q^2=Q_\mathrm{t}^2 + Q_{L,B}^2$, where
$Q_\mathrm{t}^2=(\vec{p}_\mathrm{t1}-\vec{p}_\mathrm{t2})^2$ is the component transverse to the thrust axis
and
$Q_{L,B}^2=(p_\mathrm{l1}-p_\mathrm{l2})^2-(E_1-E_2)^2$ combines the longitudinal momentum and energy differences.
Again, $R_2$ is maximal along the line $Q=0$. 
This is a non-trivial result.
For a hydrodynamical type of source,
on the contrary, BEC decrease when any of the relative momentum components is large~\cite{Csorgo:1999sj,Csorgo:1995bi}.
%

%\section{A region of anti-correlations}
%\label{sec:anti}

\begin{figure}
\includegraphics[width=0.5\textwidth,bb=40 50 519 650,clip]{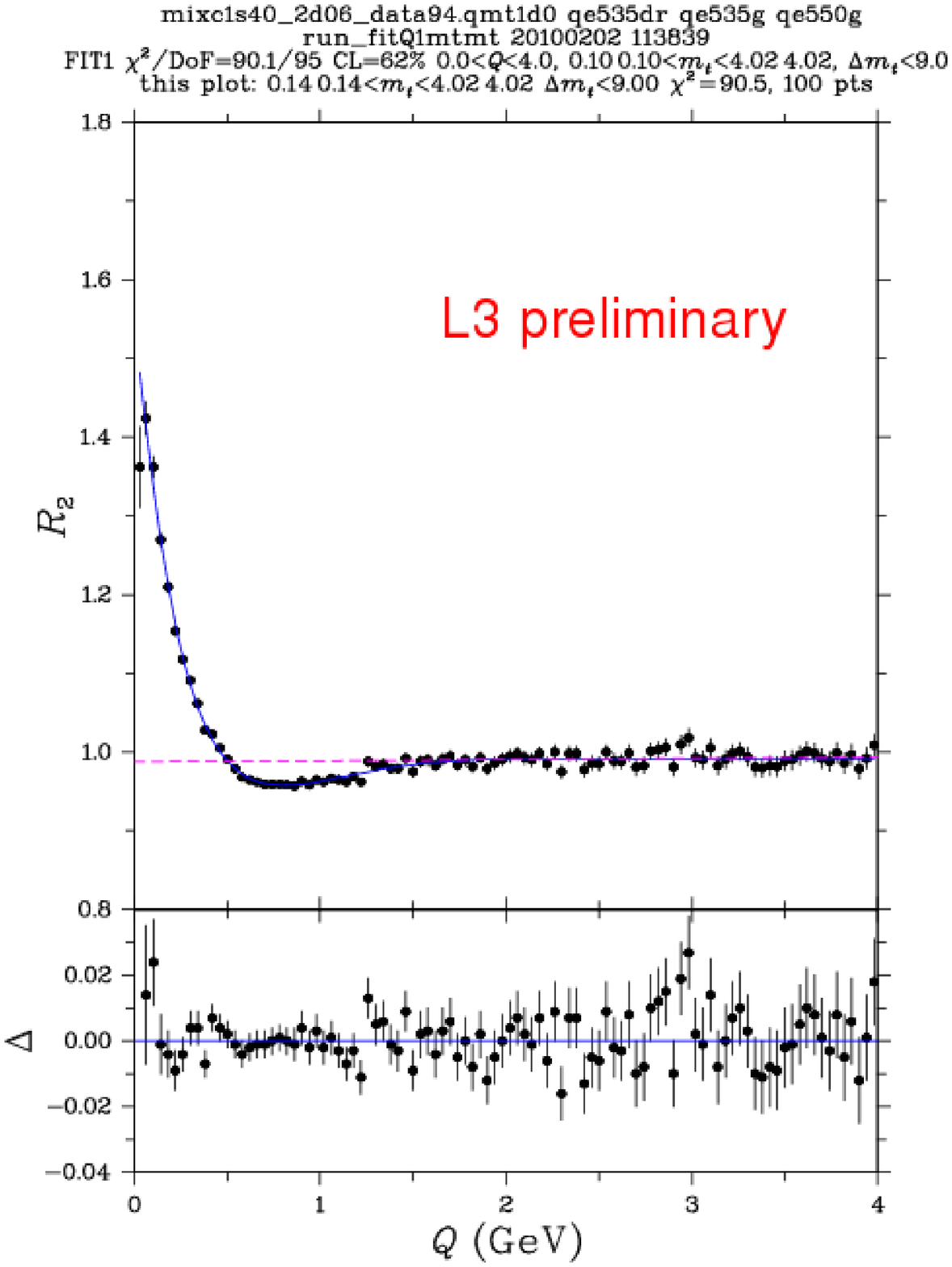}% Here is how to import EPS art
\caption{\label{f:anti}\Lthree data on Bose-Einstein correlations of charged particles are
compared to a  $\tau $ model fit. The fit
describes the peak structure at low $Q \le 0.5 $ GeV, as well as the region of anti-correlations
in $ 0.5 < Q \le 1.5$ GeV around $Q \approx 1$ GeV with a $\chi^2/NDF = 90.1/95$. Note that the background distribution in the 
$Q \ge 2$ GeV is within errors flat, all long-range correlations have been removed. 
$\Delta$ indicates theory - data.}
\end{figure}

In the region of 0 $\le Q \le$ 0.5  GeV, we observe as usual, a positive correlation, due to the
Bose-Einstein symmetrization effect of like-sign charged identical boson pairs.
In the region 0.5 $\le Q \le 1.6$ GeV,  the measured \Lthree correlation function $R_2$ 
decreases below unity~\footnote{Actually, dipping below the value of $\gamma(1+\epsilon Q)$.},
which is indicative of an anti-correlation.
This is clearly seen
in \Fig{f:anti} 
by comparing the data in this region  to an extrapolation of a linear fit, $\gamma(1+\epsilon Q)$
that is fitted to  our data in the region $Q\ge 1.6\,\GeV$, 
where $\gamma$ as an absolute normalization constant and $\epsilon$ is a measure
of long-range, residual non-Bose-Einstein correlations in our measurement. The extrapolation to the low 
values of the invariant relative momentum   $Q$ is indicated by the dashed line on \Fig{f:anti}.
Correlation functions with 1 + positive definit forms are  by definition unable to
describe this dip in $R_2$. This is the primary reason for the failure of 
several $Q$ dependent parameterizations.  We note that this dip is less apparent,
if one only plots (and fits) $R_2$ for $Q<2$ GeV as has usually been done in the past.

Many parametrizations discussed earlier, for example Gaussian or L\'evy source distributions
as well as Edgeworth expansion, have been shown before to be 
insufficient to describe the BEC~\cite{Novak:2008zz,Metzger:2008zza,Novak:2009xq}. 
These parameterizations assume a static source:  
the parameter $R$, representing the size of the source as seen in the rest frame of the
pion pair, is a time independent constant.  It has, however, been observed that $R$ depends on the transverse mass,
$\mt=\sqrt{m^2+\pt^2}=\sqrt{E^2-p_z^2}$, of the pions~\cite{collaboration:2007he}.
It has been shown~\cite{Bialas:1999hc,Bialas:2000yi} that this dependence can be understood if the produced
pions satisfy, approximately, the (generalized) Bjorken-Gottfried
condition~\cite{Gottfried:1972vv,Bjorken:1972at}, whereby
the four-momentum of a produced particle and the space-time position at which it is produced are linearly
related.
%related:
%
%\begin{equation} \label{eq:GB-corr}
%   x^\mu = d p^\mu \;.
%\end{equation}
Such a correlation between space-time and momentum-energy is  also a feature of the Lund string model,
which, incorporated in \JETSET~\cite{Sjostrand:1993yb},
is very successful in describing detailed features of the hadronic final states of \Pep\Pem\ annihilation.

A model which predicts such a $Q$-dependence while incorporating the Bjorken-Gottfried condition
is the so-called \taumodel , introduced in ref.~\cite{Csorgo:1990up}.
In this model, it is assumed that the average production point in the overall center-of-mass system,
$\overline{x}=(\overline{t},\overline{r}_x,\overline{r}_y,\overline{r}_z)$, of particles with a given
four-momentum $p=(E,p_x,p_y,p_z)$ is 
\begin{equation} \label{eq:tau-corr}
   \overline{x}^\mu (p^\mu)  = a\tau p^\mu \;.
\end{equation}
In the case of two-jet events,
  $a=1/\mt$
where
 $\mt=\sqrt{m^2+\pt^2}=\sqrt{E^2-p_z^2}$ is the transverse mass
and $\tau = \sqrt{\overline{t}^2 - \overline{r}_z^2}$ is the longitudinal proper time.
%~\footnote{The
%terminology `longitudinal' proper time and `transverse' mass seems customary in the literature
%even though their definitions are analogous $\tau = \sqrt{\overline{t}^2 - \overline{r}_z^2}$
%and                                         $ \mt = \sqrt{E^2            - p_z^2}$.}.
For isotropically distributed particle production, the transverse mass is replaced by the
mass in the definition of $a$ and $\tau$ is the proper time. 
In the case of three-jet events the relation is more complicated.
The second assumption of the \taumodel\ is that the distribution of $x^\mu (p^\mu)$ about its average,
$\delta_\Delta ( x^\mu(p^\mu) - \overline{x}^\mu (p^\mu) )$, is narrower than the
proper-time distribution, $H(\tau)$.
The emission function of the \taumodel\ is
\begin{equation} \label{eq:source}
  S(x,p) = \int_0^{\infty} \mathrm{d}\tau H(\tau)\delta_{\Delta}(x-a\tau p) \rho_1(p) \;,
\end{equation}
where $H(\tau)$ is the (longitudinal) proper-time distribution, the factor
$\delta_{\Delta}(x-a\tau p)$ describes the strength of the correlations between
coordinate space and momentum space variables and $\rho_1(p)$ is the experimentally
measurable single-particle spectrum.

The two-pion distribution, $\rho_2(p_1,p_2)$, is related to $S(x,p)$, in the plane-wave approximation,
by the Yano-Koonin formula~\cite{Yano:1978gk}.
The resulting two-particle Bose-Einstein correlation function is indeed found to depend on the invariant
relative mometum variable $Q$, as well as on the values of $a$ of the two particles\cite{Csorgo:2008ah}:
\begin{equation} \label{eq:levyR2}
   R_2(p_1,p_2) = 1 +         \mathrm{Re} \widetilde{H}\left(\frac{a_1 Q^2}{2}\right)
                                          \widetilde{H}\left(\frac{a_2 Q^2}{2}\right) \;,
\end{equation}
where $\widetilde{H}(\omega) = \int \mathrm{d} \tau H(\tau) \exp(i \omega \tau)$
is the Fourier transform (characteristic function) of $H(\tau)$.
Note that $H(\tau)$ is normalized to unity. 
This formula simplifies further if $R_2$ is measured with the restriction
     $ a_1\approx a_2\approx \bar{a} $.
In that case,
%\begin{equation}     \label{eq:levyR2a1}
%   R_2(k_1,k_2) = 1 + \lambda \, \mathrm{Re} \widetilde{H}^2
%                  \left(  \frac{\bar{a}Q^2}{2} \right) \;,
%\end{equation}
for two-jet events $R_2$ becomes
\begin{equation}     \label{eq:levy2jetR2a}
   R_2(p_1,p_2) = 1 + \lambda \, \mathrm{Re} \widetilde{H}^2
                  \left(  \frac{Q^2}{2 \overline{m}_{\mathrm{t}}} \right) \;.
\end{equation}
Thus for a given average of $a$ of the two particles, $R_2$  is found to
depend only on the invariant relative momentum $Q$.
Further, the model predicts a specific dependence on $\bar{a}$, which for two-jet events is
a specific dependence on  $\overline{m}_\mathrm{t}$
\footnote{In the initial formulation of the \taumodel\
this dependence was averaged over \cite{Csorgo:1990up} due to the lack of \mt\ dependent investigations
at that time.}.

Since there is no particle production before the onset of the collision,
$H(\tau)$ should be a  one-sided distribution.
We choose a one-sided L\'evy distribution~\cite{Csorgo:2008ah}, which has the
characteristic function~\cite{Csorgo:2003uv} (for $\alpha\ne1$)
\begin{widetext}
\begin{equation} \label{eq:levy1sidecharf}
   \widetilde{H}(\omega) = \exp\left[ -\frac{1}{2}\left(\Delta\tau|\omega|\strut\right)^\alpha
          \left( 1 -  i\, \mathrm{sign}(\omega) \tan\left(\frac{\alpha\pi}{2}\right) \strut \right)
       + i\,\omega\tau_0\right]
%                                  \;,\quad \alpha\ne1   \;,
 \; ,
\end{equation}
\end{widetext}
where the parameter $\tau_0$ is the proper time of the onset of particle production
and $\Delta \tau$ is a measure of the width of the proper-time distribution.
For the special case $\alpha=1$, see, for example, \cite{Nolan:2010}.
We have tested that parameter $\tau_0$ is within errors zero, hence we fixed it to zero.
Using this simplification, 
the  characteristic function in \Eq{eq:levyR2} yields
\begin{widetext}
\begin{equation} \label{eq:levyR2a} 
   R_2(Q,a_1,a_2) =\gamma\left\{ 1 +
      \lambda \cos\left[
\tan\left(\frac{\alpha\pi}{2}\right)\left(\frac{\Delta\tau {Q^2}}{2}\right)^{\!\alpha}\frac{a_1^\alpha+a_2^\alpha}{2} \right] 
\exp \left[-\left(\frac{\Delta\tau {Q^2}}{2}\right)^{\!\alpha}\frac{a_1^\alpha+a_2^\alpha}{2} \right]\right\}
(1 + \epsilon Q)
 \; .
\end{equation}
\end{widetext}

\begin{table}[t]
\caption{Results of the fit of \Eq{eq:levyR2a} for two-jet events.
         The first uncertainty is statistical, the second systematic.
\label{tab:2jetR2}
         }
%\centering
\begin{center}
\begin{tabular}{ l r@{$\;\pm\;$}l
               }
\hline
    parameter               & \multicolumn{2}{c }{\ }           \\
\hline
  $\lambda$                 & 0.58  & $0.03^{+0.08}_{-0.24}$    \\
  \rule{0pt}{11pt}$\alpha$  & 0.47  & $0.01^{+0.04}_{-0.02}$    \\
  $\Delta\tau$   (fm)       & 1.56  & $0.12^{+0.32}_{-0.45}$    \\
  $\epsilon$ ($\invGeV$)        & 0.001 & $0.001\pm0.003$           \\
  $\gamma$                  & 0.988 & $0.002^{+0.006}_{-0.002}$ \\
\hline
  \chisq/DoF                & \multicolumn{2}{c }{90/95}        \\
  confidence level          & \multicolumn{2}{c }{62\%}         \\
\hline
\end{tabular}
\end{center}
\end{table}

\begin{table}[t]
\caption{Confidence levels and the values of $\lambda$ found in fits of \Eq{eq:levyR2a} for two-jet events
         in various  regions of the $m_{\mathrm{t}1}$-$m_{\mathrm{t}2}$ plane
         with $\alpha$ and  $\Delta\tau$                               fixed to the result of the fit to the entire plane.
\label{tab:2jetmtCL}
         }
%\centering
\begin{center}
\begin{tabular}{cc|c|c|c}
\hline
    \multicolumn{2}{c|}{\mt\ regions ($\GeV$)}& \multicolumn{2}{c|}{average}  & confidence \\
    $m_{\mathrm{t}1}$  &  $m_{\mathrm{t}2}$ & \multicolumn{2}{c|}{\mt\ ($\GeV$)}& level      \\
                       &                    &$Q<0.4$&all&    (\%)         \\
\hline
    0.14 -- 0.26       &   0.14 -- 0.22     & 0.19 & 0.19 &  10    \\ % 11
    0.14 -- 0.34       &   0.22 -- 0.30     & 0.27 & 0.27 &  48    \\ % 12
    0.14 -- 0.46       &   0.30 -- 0.42     & 0.37 & 0.37 &  74    \\ % 13
    0.14 -- 0.66       &   0.42 -- 4.14     & 0.52 & 0.52 &  13    \\ % 14
    0.26 -- 0.42       &   0.14 -- 0.22     & 0.25 & 0.26 &  22    \\ % 21
    0.34 -- 0.46       &   0.22 -- 0.30     & 0.32 & 0.33 &  33    \\ % 22
    0.46 -- 0.58       &   0.30 -- 0.42     & 0.43 & 0.44 &  34    \\ % 23
    0.66 -- 0.86       &   0.42 -- 4.14     & 0.65 & 0.65 &  66    \\ % 24
    0.42 -- 0.62       &   0.14 -- 0.22     & 0.34 & 0.34 &  17    \\ % 31
    0.46 -- 0.70       &   0.22 -- 0.30     & 0.41 & 0.41 &  55    \\ % 32
    0.58 -- 0.82       &   0.30 -- 0.42     & 0.52 & 0.52 &  59    \\ % 33
    0.86 -- 1.22       &   0.42 -- 4.14     & 0.80 & 0.81 &  24    \\ % 34
%   0.62 -- 4.14       &   0.14 -- 0.22     & 0.48 & 0.57 &  \pho6 \\ % 41
    0.70 -- 4.14       &   0.22 -- 0.30     & 0.59 & 0.65 &  \pho4 \\ % 42
    0.82 -- 4.14       &   0.30 -- 0.42     & 0.71 & 0.76 &  11    \\ % 43
%   1.22 -- 4.14       &   0.42 -- 4.14     & 1.01 & 1.07 &  30    \\ % 44
\hline
\end{tabular}
\end{center}
\end{table}

The result of fitting \Eq{eq:levyR2a} to the \Lthree two-jet event Bose-Einstein correlation
data is presented on \Fig{f:anti}. The best values of the fit parameters and their errors 
are shown in \Tab{tab:2jetR2}.
Their first uncertainty is statistical, the second systematic.
The confidence levels are shown in \Tab{tab:2jetmtCL} for varying the
transverse mass of the particles independently. The results indicate, that the \taumodel\ 
is consistent with the \Lthree data, the fit quality is good and the model
is able to describe data  well not only the low relative momentum region,
but also  the region of $0.5 \le Q \le 1.6$ GeV, where anti-correlations are observed.
In the large relative momentum region of $Q > 1.6$  GeV, no significant long-range correlations
are found and the corresponding parameter $\epsilon$ is measured to be zero within errors.
Based on the analysis of the Bose-Einstein correlations and the single particle spectra
in terms of the \taumodel\, the space-time evolution of the particle emitting source
can also be reconstructed~\cite{Csorgo:2008ah}. 
The first \Lthree preliminary results on such an extremely fast movie
were reported in refs.~\cite{Metzger:2008zza,Novak:2009xq}.

{\it Acknowledgments:}
T. Cs. is grateful to Prof. R. Glauber for stimulating discussions and
to V. Kuvshinov and his team for organizing
an inspiring and useful meeting in Gomel.
This research was supported by the OTKA grants NK73143 and T049466,
and by a Senior Leaders and Scholars Award of HAESF,
the Hungarian-American Enterprise Scholarship Fund.

%\label{last}
\end{document}